\documentclass[fleqn,showkeys,11pt]{revtex4}
\usepackage{graphicx}
\usepackage{amssymb}
\usepackage{amsmath}
\usepackage{bm}

\usepackage{amsfonts,amssymb,amsmath}
\usepackage{graphicx}

\usepackage{graphics,hyperref}
\usepackage{amssymb}
\usepackage{amsmath}
\usepackage{color}
\usepackage{bm}
\usepackage{float}

\begin{document}

\title{Efficient Entanglement Measure for Graph States}
\author{Saeed Haddadi$^{1,2}$, Ahmad Akhound$^{2}$, and Mohammad Ali Chaman Motlagh$^{2}$}

\affiliation{$^{1}$ Faculty of Physics, Semnan University, P.O.Box 35195-363, Semnan, Iran\\$^{2}$ Department of Physics, Payame Noor University, P.O.Box 19395-3697, Tehran, Iran\\haddadi@physicist.net}

\begin{abstract}
In this paper, we study the multipartite entanglement properties of graph states up to seven qubits. Our analysis shows that the generalized concurrence measure is more efficient than geometric entanglement measure for measuring entanglement quantity in the multi-qubit graph states.
\keywords{Entanglement; Graph state; Efficient measure}
\end{abstract}

\keywords{Entanglement; Graph state; Efficient measure}

\maketitle

\section{Introduction}
\label{intro}
Entanglement is one of the most striking features of quantum mechanics, which has many applications in quantum information theory such as quantum teleportation \cite{01,02,00,Ming,03}, quantum cryptography \cite{04}, quantum dense coding \cite{05}, and quantum computing \cite{06,07,08}. Up to now, different measures have been introduced for measuring entanglement quantity in bipartite and multipartite systems, however, no one can claim which measure would be more efficient than the others yet \cite{09}. Since each measure has advantages over the others in various conditions \cite{09,10,11,111,12,vv}.

The entangled states have a fundamental role in multipartite systems for application in quantum information processing and communications as well. Graph states \cite{Hein01,Hein02} are a type of $n$-qubit pure state with various applications in quantum information processing\cite{g0,0g0,g00,g000,g1,g11,g111,g1111,g2,gx,gx2,g22,g222,g2222,g3,g33,g4,g5,g6,g7}. An $n$-qubit graph state $|G\rangle$ is a pure state associated to a graph $G=(V,E)$. The graph $G=(V,E)$ gives both a recipe for preparing $|G\rangle$ and mathematical characterization of $|G\rangle$.
Graph states play several significant roles in quantum information theory, e.g, in quantum computation, quantum error-correction, quantum simulation, entanglement distillation protocols, multipartite purification schemes, GHZ or all-versus-nothing proofs of Bell's theorem.
Graph states are essential for quantum communication protocols, including teleportation, entanglement-based quantum key distribution, secret sharing, and reduction of communication complexity. The graph state formalism is a useful concept which authorizes a detailed classification of $n$-qubit entanglement.

The authors\cite{Nest03} detected that the serial application of a transformation is sufficient to generate the complete equivalence class of graph states under local unitary (LU) operations within the Clifford group; this simple transformation is Local Complementation (LC). All local complimentary (LC) equivalent graph states have the same values of entanglement, accordingly only LC inequivalent graph states should be considered concerning with the entanglement\cite{Hein01,Hein02,wang}. The number of non-isomorphic and non-LC-equivalent connected graphs up to seven qubits is 45, which we here calculate their entanglement by two useful entanglement criteria.

So far, many papers have been published on graph states and hypergraph states in which the authors used geometric entanglement measure (GEM) for measuring the amount of entanglement \cite{s1,s2,s3,s4,s5,s6,s7}. In this work, we demonstrate that not only the GEM does not an efficient entanglement measure, but also it is actually an unsuitable measure for measuring entanglement quantity of multi-qubit graph states. This work is structured as follows: Sec. \ref{sec2} is dedicated to the expression of basic concepts about the graph states. In Sec. \ref{sec3}, we present the generalized concurrence measure (GCM) and GEM which are used in this work as multipartite entanglement measures.  Sec. \ref{sec4} is devoted to the classification of graph states up to seven qubits under non-isomorphic and non-LC-equivalent connected graphs and multipartite entanglement measures. Finally, we conclude and summarize in the last section.

\section{Graph States and Definitions}\label{sec2}
An $n$-qubit graph state corresponding to the graph $ G=(V,E) $ is given by the following equation\cite{Hein01,Hein02}
\begin{equation}
\label{E1}
 |G\rangle\:=\prod_{\{i, j\}\in E}CZ_{ij}|+\rangle_{x}^{\otimes n},
\end{equation}
where $ |+\rangle_{x}=\frac{1}{\sqrt{2}}\{|0\rangle + |1\rangle \} $ with basis vectors $|0\rangle\doteq\binom{1}{0}$ and  $|1\rangle\doteq\binom{0}{1}$ is an eigenstate (eigenvector) of the Pauli matrix $ \sigma_{x} $ with eigenvalue +1. Now, for each edge connecting two qubits, $i$ and $j$, it is applied the controlled-$Z$ ($CZ$) gate between qubits $i$ and $j$

\begin{equation}\label{E2}
CZ =|00\rangle\langle00|+|01\rangle\langle01|+|10\rangle\langle10|-|11\rangle\langle11|.
\end{equation}

In each $ n $-qubit system, the number of simple graph states is $ 2^{\binom{n}{2}} $. So for a 3-qubit system, there are 8 graph states, for 4-qubit system there are 64 graph states etc., which their non-isomorphic and non-LC-equivalent connected graphs proposed by Hein et al. are plotted in Ref. \cite{Hein01,Hein02}. There are: an inequivalent 2-qubit graph, an inequivalent 3-qubit graph, two 4-qubit graphs (No. 3, No. 4), four 5-qubit graphs (No. 5 - No. 8), eleven 6-qubit graphs (No. 9 - No. 19), twenty six 7-qubit graphs (No. 20 - No. 45).
Therefore, many graphs can be converted to each other by LU transformations or by permutations of the vertices. Consequently, they have the same value of entanglement\cite{s5}.

\vspace*{4mm}

\textbf{Definition 2.1 (graph isomorphism).}
Two graphs $ G_{1}=(V_{1},E_{1}) $ and $ G_{2}=(V_{2},E_{2}) $ are called isomorphic $ (G_{1}\cong G_{2}) $ if there is a bijection $ f: V_{1}\rightarrow V_{2} $ is a mapping of a graph onto itself  between a set of vertices such that $ \{a,b\}\in E_{1} $ if and only if (iff) $ \{f(a),f(b)\}\in E_{2} $ \cite{k1,West04,Bondy05}.

\vspace*{4mm}

\textbf{Definition 2.2 (local complementation).}
By local complementation (LC) of a graph $ G=(V,E) $ at some vertex of $ a\in V $ can obtain an LC-equivalent graph state as $|\tau_{a}(G)\rangle=U_{a}^{\tau}(G)|G\rangle,$ in which
\begin{equation}
U_{a}^{\tau}(G)=\exp(-i\frac{\pi}{4}\sigma_{x}^{a})\prod_{b\in N_{a}}\exp(i\frac{\pi}{4}\sigma_{z}^{b}),
\end{equation}
is a local Clifford unitary\cite{Hein01,Hein02}. Also, $ N_{a}$ is neighbors of vertex $a$ and $\sigma_{\alpha}(\alpha=x, z)$ are the Pauli matrices. Hence, two graph states $ |G\rangle $ and $ |G^{\prime}\rangle $ are LC-equivalent iff the corresponding graphs are linked by a sequence of local complementations. In the other words, LC centered on a qubit $a$ is visualized easily as a transformation of the subgraph of the $a$th qubit’s neighbors, such that one edge between two neighbors of $a$ is removed if the two neighbors are themselves connected, or one edge is added otherwise. \cite{s2}.

\section{Multipartite Entanglement Measures}\label{sec3}
In order to investigate the entanglement depth of the graph states, we first review the entanglement measures. For multipartite systems, several measures of entanglement have been proposed \cite{09}. Here, we use two important multipartite entanglement measures for measuring entanglement quantity of graph states, namely GCM and GEM. First, for an $n$-partite pure state $|\psi\rangle$, the GCM is defined as \cite{Carvalho06,Mintert07,Fei08,Fei09}

\begin{equation}
\label{E3}
\textmd{GCM}(|\psi\rangle)=2^{1-{\frac{n}{2}}}\times\left(2^{n}-2-\sum_{\alpha}\textmd{Tr}\rho_{\alpha}^{2}\right)^{\frac{1}{2}},
\end{equation}
where $ \alpha $ labels as all different subsystems of the $ n $-partite system and $ \rho_{\alpha} $ are the corresponding reduced density matrices which determined by taking the partial trace of $ \hat{\rho}=|\psi\rangle\langle\psi| $. For example, for a 3-qubit system, one requires to determine the reduced density matrices $\rho_{1}=Tr_{23}\hat{\rho}$, $\rho_{2}=Tr_{13}\hat{\rho}$, $\rho_{3}=Tr_{12}\hat{\rho}$, $\rho_{12}=Tr_{3}\hat{\rho}$, $\rho_{13}=Tr_{2}\hat{\rho}$, and $\rho_{23}=Tr_{1}\hat{\rho}$.
Second, let us consider GEM, which is defined for pure states as \cite{Wei1,Wei2,Wei3,Wei4,Hajdusek11,Zhang11}
\begin{equation}
\label{E4}
\textmd{GEM}(|\psi\rangle) = 1- \max_{|\phi\rangle=|a\rangle|b\rangle|c\rangle}|\langle\phi|\psi\rangle|^{2},
\end{equation}
where $|\phi\rangle=|a\rangle|b\rangle|c\rangle$ is the set of product states. This is the distance between the product state $|\phi\rangle$ and another state as $|\psi\rangle$ in terms of fidelity $F_{\phi}=|\langle\phi|\psi\rangle|^{2}$. For mixed states, this entanglement monotone can be extended using the convex roof.

\section{Efficient Entanglement Measure}\label{sec4}
We note that up to seven qubits, there are 45 classes of graph states which are not equivalent under one-qubit unitary transformations \cite{Hein01,Hein02}. We first obtain 45 graph states corresponding to 45 graphs which are all plotted in Ref. \cite{Hein01,Hein02} and numbered. However, to save space, we only address some of those. Explicitly,

\begin{align}
\begin{split}
|G_{1}\rangle&=\frac{1}{\sqrt{2}}\{|0\rangle_{1}|+\rangle_{2}+|1\rangle_{1}|-\rangle_{2}\},\nonumber\\
|G_{2}\rangle&=\frac{1}{\sqrt{2}}\{|0\rangle_{1}|+\rangle_{2}|+\rangle_{3}+|1\rangle_{1}|-\rangle_{2}|-\rangle_{3}\},\\
|G_{3}\rangle&=\frac{1}{\sqrt{2}}\{|0\rangle_{1}|+\rangle_{2}|+\rangle_{3}|+\rangle_{4}+|1\rangle_{1}|-\rangle_{2}|-\rangle_{3}|-\rangle_{4}\},\\
|G_{5}\rangle&=\frac{1}{\sqrt{2}}\{|0\rangle_{1}|+\rangle_{2}|+\rangle_{3}|+\rangle_{4}|+\rangle_{5}+|1\rangle_{1}|-\rangle_{2}|-\rangle_{3}|-\rangle_{4}|-\rangle_{5}\},\\
|G_{7}\rangle&=\frac{1}{2}\{|+\rangle_{1}|0\rangle_{2}|+\rangle_{3}|0\rangle_{4}|+\rangle_{5}+|+\rangle_{1}|0\rangle_{2}|-\rangle_{3}|1\rangle_{4}|-\rangle_{5}\\
&\quad\;+|-\rangle_{1}|1\rangle_{2}|-\rangle_{3}|0\rangle_{4}|+\rangle_{5}+|-\rangle_{1}|1\rangle_{2}|+\rangle_{3}|1\rangle_{4}|-\rangle_{5}\},\\
|G_{12}\rangle&=\frac{1}{2}\{|+\rangle_{1}|0\rangle_{2}|+\rangle_{3}|0\rangle_{4}|+\rangle_{5}|+\rangle_{6}+|+\rangle_{1}|0\rangle_{2}|-\rangle_{3}|1\rangle_{4}|-\rangle_{5}|+\rangle_{6}\\
&\quad\;+|-\rangle_{1}|1\rangle_{2}|-\rangle_{3}|0\rangle_{4}|+\rangle_{5}|-\rangle_{6}+|-\rangle_{1}|1\rangle_{2}|+\rangle_{3}|1\rangle_{4}|-\rangle_{5}|-\rangle_{6}\},\\
|G_{15}\rangle&=\frac{1}{2}\{|+\rangle_{1}|+\rangle_{2}|+\rangle_{3}|0\rangle_{4}|+\rangle_{5}|0\rangle_{6}+|-\rangle_{1}|+\rangle_{2}|-\rangle_{3}|0\rangle_{4}|-\rangle_{5}|1\rangle_{6}\\
&\quad\;+|+\rangle_{1}|-\rangle_{2}|+\rangle_{3}|1\rangle_{4}|-\rangle_{5}|1\rangle_{6}+|-\rangle_{1}|-\rangle_{2}|-\rangle_{3}|1\rangle_{4}|+\rangle_{5}|0\rangle_{6}\},\\
|G_{20}\rangle&=\frac{1}{\sqrt{2}}\{|0\rangle_{1}|+\rangle_{2}|+\rangle_{3}|+\rangle_{4}|+\rangle_{5}|+\rangle_{6}|+\rangle_{7}+|1\rangle_{1}|-\rangle_{2}|-\rangle_{3}|-\rangle_{4}|-\rangle_{5}|-\rangle_{6}|-\rangle_{7}\},\\
\end{split}
\end{align}

where $ |+\rangle_{x}=\frac{1}{\sqrt{2}}\{|0\rangle + |1\rangle \}$ and  $ |-\rangle_{x}=\frac{1}{\sqrt{2}}\{|0\rangle - |1\rangle \}$ are eigenstates of the Pauli matrix $ \sigma_{x} $ with eigenvalues $\pm1$.
For example, in order to calculate the amount of entanglement in 3-qubit graph state (No. 2) by GCM and GEM, we obtain by direct numerical optimization GCM($|G_{2}\rangle$)=1.22474, and GEM($|G_{2}\rangle$)=0.5 with the edges $E_{2}=\{\{1, 2\}, \{1, 3\}\}$. There are 45 set of edges corresponding to these graphs which are reported in appendix \ref{sec:app}.

Next, we apply two entanglement measures (see Eqs. (\ref{E3}) and (\ref{E4})) for all 45 graph states and we enter numerical values in Table \ref{tab:1} and \ref{tab:2}. Hence, we have classified the 45 graphs into 27 and 7 classes, according to the GCM and GEM, respectively.

\begin{table}[H]
\centering
\caption{The classification of non-isomorphic and non-LC-equivalent connected graphs up to seven qubits based on generalized concurrence measure (GCM).}
\label{tab:1}       
\begin{tabular}{l c c| c c c}
\hline\noalign{\smallskip}
Class &\qquad GCM  &\qquad Graph No. &\qquad Class &\qquad GCM &\qquad Graph No.  \\

  \hline
  1 &\qquad 1 &\qquad 1 &\qquad 15 &\qquad 1.63936 &\qquad 14, 16\\
  2 &\qquad 1.22474 &\qquad 2 &\qquad 16 &\qquad 1.64886 &\qquad 24\\
  3 &\qquad 1.32288 &\qquad 3 &\qquad 17 &\qquad 1.65831 &\qquad 17, 25, 31\\
  4 &\qquad 1.36931 &\qquad 5 &\qquad 18 &\qquad 1.67705 &\qquad 18, 26\\
  5 &\qquad 1.39194 &\qquad 9 &\qquad 19 &\qquad 1.68634 &\qquad 27, 32\\
  6 &\qquad 1.40312 &\qquad 20 &\qquad 20 &\qquad 1.69558 &\qquad 19, 28, 33\\
  7 &\qquad 1.41421 &\qquad 4 &\qquad 21 &\qquad 1.70477 &\qquad 29, 34\\
  8 &\qquad 1.50000 &\qquad 6 &\qquad 22 &\qquad 1.71391 &\qquad 30, 35, 36\\
  9 &\qquad 1.54110 &\qquad 7, 10 &\qquad 23 &\qquad 1.72301 &\qquad 37\\
  10 &\qquad 1.56125 &\qquad 21 &\qquad 24 &\qquad 1.73205 &\qquad 38, 39\\
  11 &\qquad 1.58114 &\qquad 8, 11 &\qquad 25 &\qquad 1.74105 &\qquad 41\\
  12 &\qquad 1.60078 &\qquad 12 &\qquad 26 &\qquad 1.75000 &\qquad 40, 42, 43, 45\\
  13 &\qquad 1.62019 &\qquad 13, 15, 22 &\qquad 27 &\qquad 1.75891 &\qquad 44\\
  14 &\qquad 1.62980 &\qquad 23 &\qquad  &\qquad  &\qquad \\
  \hline
\end{tabular}
\end{table}

\begin{table}[H]
\centering
\caption{The classification of non-isomorphic and non-LC-equivalent connected graphs up to seven qubits based on geometric entanglement measure (GEM).}
\label{tab:2}       
\begin{tabular}{l c c}
\hline\noalign{\smallskip}
Class &\qquad GEM  &\qquad Graph No.   \\

  \hline
  1 &\qquad 0.50000 &\qquad 1, 2, 3, 5, 9, 20 \\
  2 &\qquad 0.75000 &\qquad 4, 6, 7, 10-12, 15, 21-24, 31 \\
  3 &\qquad 0.86855 &\qquad 8 \\
  4 &\qquad 0.87500 &\qquad 13-14, 16-18, 25-30, 32-38, 43 \\
  5 &\qquad 0.91667 &\qquad 19 \\
  6 &\qquad 0.93428 &\qquad 39, 41, 45 \\
  7 &\qquad 0.93750 &\qquad 40, 42, 44 \\
  \hline
\end{tabular}
\end{table}

In the last step, we collect the classification results in Table \ref{tab:3}. We here should comment about the results in the tables. The resolution power (RP) in Table \ref{tab:3} is computed using the following equation
\begin{equation}
\label{E5}
\textmd{RP}=\frac{\eta[\chi]}{\eta[\kappa]}\times100,
\end{equation}

where $\eta[\chi]$ denotes the number of classes based on GCM or GEM and $\eta[\kappa]$ is the number of categories according to non-isomorphic and non-LC-equivalent connected graphs. Therefore, the RP in Table \ref{tab:3} is the ratio of the number of classifications taken from multipartite entanglement measures (see Eqs. (\ref{E3}) and (\ref{E4})) to the number of non-isomorphic and non-LC-equivalent connected graphs. Using the method of comparing the results of these classes with classification of graph states under non-isomorphic and non-LC-equivalent connected graphs, as we call RP method, we found that the GCM seems more efficient than GEM for multi-qubit graph states.

\begin{table}[H]
\centering
\caption{Classification of graph states up to seven qubits based on GCM and GEM. In the second column, the number of graph states with $n$ =2, 3, 4, 5, 6 and 7 vertices is listed which are categorized by generalized concurrence measure ($\eta$ [GCM]). In the third column, the number of graph states which are categorized by geometric entanglement measure ($\eta$ [GEM]). The values in the fourth column (number of non-isomorphic and non-LC-equivalent connected graphs) were computed in Refs. \cite{Hein02} and \cite{Danielsen12} together with a database of representatives for each equivalence class ($\eta [\kappa]$). In the fifth column and the last column give the results of the resolution power (RP).}
\label{tab:3}       
\begin{tabular}{l c c c c c}
\hline\noalign{\smallskip}
n &\qquad $\eta$ [GCM]  &\qquad $\eta$ [GEM] &\qquad $\eta [\kappa]$ &\qquad RP [GCM] &\qquad RP [GEM]  \\

  \hline
  2 &\qquad 1 &\qquad 1 &\qquad 1 &\qquad 100\% &\qquad 100\%\\
  3 &\qquad 1 &\qquad 1 &\qquad 1 &\qquad 100\% &\qquad 100\%\\
  4 &\qquad 2 &\qquad 2 &\qquad 2 &\qquad 100\% &\qquad 100\%\\
  5 &\qquad 4 &\qquad 3 &\qquad 4 &\qquad 100\% &\qquad 75\%\\
  6 &\qquad 9 &\qquad 4 &\qquad 11 &\qquad 81.82\% &\qquad 36.36\%\\ \vspace*{1mm}
  7 &\qquad 16 &\qquad 5 &\qquad 26 &\qquad 61.54\% &\qquad 19.23\%\\
  Up to 7 &\qquad 27 &\qquad 7 &\qquad 45 &\qquad 60\% &\qquad 15.55\%\\
  \hline
\end{tabular}
\end{table}

\section{Conclusion}
We propose two novel classifications for the entanglement in graph states up to seven qubits based on generalized concurrence measure (GCM) and geometric entanglement measure (GEM), and we also compare those with the classification under non-isomorphic and non-LC-equivalent connected graphs. Numerical values reveal that the 45 graph states are classified according to the GCM and GEM into 27 and 7 categories, respectively. Accordingly, our results propose that the GCM is a suitable entanglement measure for measuring entanglement quantity of multi-qubit graph states, and it is also more efficient than GEM. The suggested approach (RP method) can be employed to recognize the proper performance of each new measure proposed for measuring entanglement in multi-qubit graph states. We believe that the investigation of graph states is necessary for a better understanding of multipartite systems and validation of multipartite entanglement measures.

\vspace*{40mm}

\appendix

\section{Appendix}
\label{sec:app}

In this appendix, we give the set of edges corresponding to the non-isomorphic and non-LC-equivalent connected graphs up to seven qubits.

\begin{table}[H]
\centering
\caption{The set of edges corresponding to the non-isomorphic and non-LC-equivalent connected graphs.}
\label{tab:4}       
\scalebox{0.7}{
\begin{tabular}{l l}
\hline\noalign{\smallskip}
Graph No. &\qquad\qquad\qquad\qquad\qquad\qquad\qquad Edges   \\

  \hline
  1 &\qquad \{\{1, 2\}\} \\
  2 &\qquad \{\{1, 2\}, \{1, 3\}\} \\
  3 &\qquad \{\{1, 2\}, \{1, 3\}, \{1, 4\}\} \\
  4 &\qquad \{\{1, 2\}, \{2, 3\}, \{3, 4\}\} \\
  5 &\qquad \{\{1, 2\}, \{1, 3\}, \{1, 4\}, \{1, 5\}\} \\
  6 &\qquad \{\{1, 2\}, \{2, 3\}, \{3, 4\}, \{2, 5\}\} \\
  7 &\qquad \{\{1, 2\}, \{2, 3\}, \{3, 4\}, \{4, 5\}\} \\
  8 &\qquad \{\{1, 2\}, \{2, 3\}, \{3, 4\}, \{4, 5\}, \{1, 5\}\} \\
  9 &\qquad \{\{1, 2\}, \{1, 3\}, \{1, 4\}, \{1, 5\}, \{1, 6\}\} \\
  10 &\qquad \{\{1, 6\}, \{2, 6\}, \{3, 6\}, \{4, 5\}, \{5, 6\}\} \\
  11 &\qquad \{\{1, 6\}, \{2, 6\}, \{3, 5\}, \{4, 5\}, \{5, 6\}\} \\
  12 &\qquad \{\{1, 2\}, \{2, 3\}, \{3, 4\}, \{4, 5\}, \{2, 6\}\} \\
  13 &\qquad \{\{1, 2\}, \{2, 3\}, \{3, 4\}, \{4, 5\}, \{3, 6\}\} \\
  14 &\qquad \{\{1, 2\}, \{2, 3\}, \{3, 4\}, \{4, 5\}, \{5, 6\}\} \\
  15 &\qquad \{\{1, 6\}, \{2, 4\}, \{3, 4\}, \{4, 5\}, \{5, 6\}, \{3, 6\}\} \\
  16 &\qquad \{\{1, 2\}, \{2, 3\}, \{3, 4\}, \{4, 5\}, \{2, 4\}, \{3, 6\}\} \\
  17 &\qquad \{\{1, 2\}, \{2, 3\}, \{3, 4\}, \{4, 5\}, \{5, 1\}, \{1, 6\}\} \\
  18 &\qquad \{\{1, 2\}, \{2, 3\}, \{3, 4\}, \{4, 5\}, \{5, 6\}, \{1, 6\}\} \\
  19 &\qquad \{\{1, 2\}, \{2, 3\}, \{3, 4\}, \{4, 5\}, \{5, 6\}, \{1, 6\}, \{1, 3\}, \{4, 6\}, \{2, 5\}\} \\
  20 &\qquad \{\{1, 2\}, \{1, 3\}, \{1, 4\}, \{1, 5\}, \{1, 6\}, \{1, 7\}\} \\
  21 &\qquad \{\{1, 7\}, \{2, 7\}, \{3, 7\}, \{4, 7\}, \{5, 6\}, \{6, 7\}\} \\
  22 &\qquad \{\{1, 7\}, \{2, 7\}, \{3, 7\}, \{4, 6\}, \{5, 6\}, \{6, 7\}\} \\
  23 &\qquad \{\{1, 7\}, \{2, 7\}, \{3, 7\}, \{4, 5\}, \{5, 6\}, \{6, 7\}\} \\
  24 &\qquad \{\{1, 7\}, \{2, 7\}, \{3, 5\}, \{4, 5\}, \{5, 6\}, \{6, 7\}\} \\
  25 &\qquad \{\{1, 2\}, \{1, 7\}, \{3, 7\}, \{4, 7\}, \{5, 6\}, \{6, 7\}\} \\
  26 &\qquad \{\{1, 7\}, \{2, 7\}, \{3, 6\}, \{4, 5\}, \{5, 6\}, \{6, 7\}\} \\
  27 &\qquad \{\{1, 2\}, \{2, 7\}, \{2, 3\}, \{4, 3\}, \{5, 4\}, \{6, 5\}\} \\
  28 &\qquad \{\{1, 2\}, \{2, 3\}, \{3, 4\}, \{3, 5\}, \{5, 6\}, \{6, 7\}\} \\
  29 &\qquad \{\{1, 2\}, \{2, 3\}, \{3, 4\}, \{4, 5\}, \{3, 6\}, \{6, 7\}\} \\
  30 &\qquad \{\{1, 2\}, \{2, 3\}, \{3, 4\}, \{4, 5\}, \{5, 6\}, \{6, 7\}\} \\
  31 &\qquad \{\{2, 3\}, \{3, 4\}, \{4, 5\}, \{5, 6\}, \{5, 7\}, \{1, 3\}, \{6, 3\}\} \\
  32 &\qquad \{\{1, 7\}, \{2, 7\}, \{3, 6\}, \{4, 5\}, \{5, 6\}, \{6, 7\}, \{5, 7\}\} \\
  33 &\qquad \{\{2, 3\}, \{3, 4\}, \{4, 5\}, \{6, 5\}, \{7, 6\}, \{3, 7\}, \{1, 3\}\} \\
  34 &\qquad \{\{2, 3\}, \{3, 4\}, \{4, 5\}, \{6, 5\}, \{7, 6\}, \{3, 6\}, \{1, 4\}\} \\
  35 &\qquad \{\{2, 3\}, \{3, 4\}, \{4, 5\}, \{6, 5\}, \{7, 6\}, \{3, 7\}, \{1, 6\}\} \\
  36 &\qquad \{\{1, 2\}, \{2, 3\}, \{3, 4\}, \{4, 5\}, \{5, 6\}, \{4, 7\}, \{3, 5\}\} \\
  37 &\qquad \{\{1, 7\}, \{2, 3\}, \{3, 4\}, \{4, 5\}, \{5, 6\}, \{6, 7\}, \{3, 7\}\} \\
  38 &\qquad \{\{1, 2\}, \{2, 3\}, \{3, 4\}, \{4, 5\}, \{5, 6\}, \{6, 7\}, \{1, 6\}\} \\
  39 &\qquad \{\{1, 2\}, \{2, 3\}, \{3, 4\}, \{4, 5\}, \{5, 6\}, \{6, 7\}, \{1, 5\}\} \\
  40 &\qquad \{\{1, 2\}, \{2, 3\}, \{3, 4\}, \{4, 5\}, \{5, 6\}, \{6, 7\}, \{1, 7\}\} \\
  41 &\qquad \{\{1, 2\}, \{2, 3\}, \{3, 4\}, \{4, 5\}, \{5, 6\}, \{6, 7\}, \{1, 5\}, \{1, 6\}\} \\
  42 &\qquad \{\{1, 3\}, \{2, 3\}, \{3, 4\}, \{4, 5\}, \{5, 6\}, \{6, 7\}, \{1, 7\}, \{2, 6\}\} \\
  43 &\qquad \{\{1, 2\}, \{2, 3\}, \{3, 4\}, \{4, 5\}, \{5, 6\}, \{5, 7\}, \{1, 4\}, \{3, 6\}\} \\
  44 &\qquad \{\{1, 4\}, \{2, 3\}, \{3, 4\}, \{4, 5\}, \{5, 6\}, \{6, 7\}, \{1, 7\}, \{2, 7\}, \{3, 5\}\} \\
  45 &\qquad \{\{1, 2\}, \{2, 3\}, \{3, 4\}, \{4, 5\}, \{5, 6\}, \{6, 7\}, \{3, 7\}, \{2, 7\}, \{2, 5\}, \{4, 6\}\} \\

  \hline
\end{tabular}}
\end{table}

\end{document}